\documentclass[prb,preprint,amsmath,amssymb,superscriptaddress,notitlepage]{revtex4-1}
\usepackage{epsfig}
\usepackage{graphicx}
\usepackage{color}
\usepackage{bm}
\usepackage{siunitx}

\def\Mn		  {{Mn$^{2+}$\ }}

\newcommand{\mr}[1]{\mathrm{#1}}
\newcommand{\unit}[1]{\,\mathrm{#1}}

\newcommand{\Ohm}{\Omega}

\newcommand{\Rxy}{R_\mr{xy}}
\newcommand{\BNV}{B_\mr{NV}}
\newcommand{\Brms}{B_\mr{rms}}
\newcommand{\Ms}{m_\mr{s}}

\begin{document}

\title{Current-induced fragmentation of antiferromagnetic domains}
\author{M.~S.~W\"{o}rnle}
\affiliation{Department of Materials, ETH Zurich, 8093 Zurich, Switzerland}
\affiliation{Department of Physics, ETH Zurich, 8093 Zurich, Switzerland}
\author{P. Welter}
\affiliation{Department of Physics, ETH Zurich, 8093 Zurich, Switzerland}
\author{Z.~Ka\v{s}par}
\affiliation{Institute of Physics, Czech Academy of Sciences, Cukrovarnick\'a 10, 162 00, Praha 6, Czech Republic}
\affiliation{Faculty of Mathematics and Physics, Charles University in Prague, Ke Karlovu 3, 121 16 Prague 2, Czech Republic}
\author{K.~Olejn\'ik}
\affiliation{Institute of Physics, Czech Academy of Sciences, Cukrovarnick\'a 10, 162 00, Praha 6, Czech Republic}
\author{V.~Nov\'ak}
\affiliation{Institute of Physics, Czech Academy of Sciences, Cukrovarnick\'a 10, 162 00, Praha 6, Czech Republic}
\author{R.~P.~Campion}
\affiliation{School of Physics and Astronomy, University of Nottingham, Nottingham NG7 2RD, United Kingdom}
\author{P.~Wadley}
\affiliation{School of Physics and Astronomy, University of Nottingham, Nottingham NG7 2RD, United Kingdom}
\author{T.~Jungwirth}
\affiliation{Institute of Physics, Czech Academy of Sciences, Cukrovarnick\'a 10, 162 00, Praha 6, Czech Republic}
\affiliation{School of Physics and Astronomy, University of Nottingham, Nottingham NG7 2RD, United Kingdom}
\author{C.~L.~Degen}
\affiliation{Department of Physics, ETH Zurich, 8093 Zurich, Switzerland}
\author{P.~Gambardella}
\affiliation{Department of Materials, ETH Zurich, 8093 Zurich, Switzerland}

\begin{abstract}
{\bf
%
%
%
Electrical and optical pulsing allow for manipulating the order parameter and magnetoresistance of antiferromagnets, opening novel prospects for digital and analog data storage in spintronic devices. Recent experiments in CuMnAs have demonstrated giant resistive switching signals in single-layer antiferromagnetic films together with analog switching and relaxation characteristics relevant for neuromorphic computing. Here we report simultaneous electrical pulsing and scanning NV magnetometry of antiferromagnetic domains in CuMnAs performed using a pump-probe scheme. We observe a nano-scale fragmentation of the antiferromagnetic domains, which is controlled by the current amplitude and independent on the current direction. The fragmented antiferromagnetic state conserves a memory of the pristine domain pattern, towards which it relaxes. Domain fragmentation coexists with permanent switching due to the reorientation of the antiferromagnetic moments. Our simultaneous imaging and resistance measurements show a correlation between the antiferromagnetic domain fragmentation and the largest resistive switching signals in CuMnAs.
}
\end{abstract}

\date{\today}
\maketitle

Antiferromagnets have been established as promising candidate materials for memory devices capable of electrical or optical writing and readout\cite{Jungwirth2018,Baltz2018,Manchon2018}. Following in the footsteps of ferromagnets, initial studies have focused on encoding information in the orientation of the magnetic order parameter, that is, in the orientation of the N\'eel vector \cite{Wadley2016,Meinert2017,Bodnar2018}. In these early studies, an antiferromagnetic variant of the current-induced spin-orbit torque \cite{Zelezny2014,Manchon2018} has been considered as the underlying writing mechanism in, for example, thin metallic films of collinear room-temperature antiferromagnets CuMnAs or Mn$_2$Au \cite{Wadley2016,Meinert2017,Bodnar2018}. Electrical 90$^{\circ}$ switching of the N\'eel vector in antiferromagnetic domains consistent with this scenario and controlled by the sign or orientation of the writing current has been confirmed by x-ray magnetic linear dichroism microscopy (XMLD-PEEM) \cite{Wadley2016,Grzybowski2017,Wadley2018,Bodnar2019,Baldrati2019}.

Electrical readout signals observed after bi(multi)-polar writing current pulses have been attributed to the anisotropic magnetoresistance (AMR) following the 90$^{\circ}$ reorientation of the N\'eel vector in the antiferromagnetic domains. Experiments with CuMnAs have shown that the onset of the AMR switching signal is correlated with the onset of the 90$^{\circ}$ domain switching seen in XMLD-PEEM at comparable threshold writing currents \cite{Grzybowski2017,Wadley2018}. In these measurements, which used 50-ms-long writing pulses with current densities of $\sim 10^6$~A~cm$^{-2}$, the relative change in resistance due to the switching was limited to $\sim$0.1\%.

Subsequent studies in CuMnAs \cite{Olejnik2017,Olejnik2018} have demonstrated switching pulse durations from ms to $\sim 1$~ps with corresponding switching current densities increasing from $\sim 10^7$~A~cm$^{-2}$ to $\sim 10^9$~A~cm$^{-2}$. Recently, the pulse duration has been reduced to 100~fs by replacing electrical switching with infrared laser pulses \cite{Kaspar2019}. The observed resistive switching signals were independent of the light polarization and could be controlled by the light intensity. The same study demonstrated that reversible switching signals could also be controlled by the amplitude of the electrical writing pulses without changing the current direction. The resulting resistive switching ratios have approached $20$\% at room temperature and $\sim 100$\% at 30~K \cite{Kaspar2019}. These exceed by $2-3$~orders of magnitude the AMR signals associated with the N\'eel vector reorientation \cite{Grzybowski2017,Wadley2018,Wang2019} and, together with the observed optical switching, point to a complementary mechanism for writing information in antiferromagnets.

In this work, we use scanning nitrogen-vacancy centre (NV) magnetometry to investigate the magnetic stray field emanating from antiferromagnetic domain textures during and after relaxation following the injection of current pulses into CuMnAs microdevices. Our study shows that the large resistive changes are correlated with a nano-scale fragmentation of domains induced by the writing current pulses. By imaging the current density distribution in microdevices with a cross geometry, we further show that the current-induced changes of the domain pattern are non-uniform across the devices. Images of the magnetic stray field acquired at varying delay times reveal that the fragmented domain patterns maintain a memory of the pristine state towards which they relax. In polarity-dependent switching experiments we observe a coexistence of fragmentation with a 180$^{\circ}$ N\'eel vector reversal in the domains. Our measurements shed light on the microscopic mechanisms leading to the electrical switching of metallic antiferromagnets and point out directions for future research in the field of antiferromagnetic spintronics.

%

\subsection*{Imaging an in-plane antiferromagnet with scanning NV magnetometry}
We investigate the antiferromagnetic domain pattern of the CuMnAs films by recording their nanoscale magnetic stray field using scanning NV magnetometry \cite{Degen2008,Balasubramanian2008}. This is a powerful microscopy technique for investigating weak magnetic patterns with high spatial resolution, with applications to nanometer-scale magnetism, superconductivity, and the imaging of current distributions \cite{Rondin2014,Casola2018}. Previous studies have demonstrated magnetic contrast in antiferromagnets with out-of-plane spin alignment such as Cr$_2$O$_3$ (Refs.~\onlinecite{Kosub2017,Appel2019a}) and the spin cycloid of multiferroic bismuth ferrite ($\text{BiFeO}_3$, Ref.~\onlinecite{Gross2017a}). A similar stray field is also expected for in-plane antiferromagnets; to our knowledge, however, such a stray field has not been measured to date. In the following, we demonstrate that scanning NV magnetometry is capable of imaging the magnetic state of in-plane antiferromagnets and we develop a model that relates the magnetic stray field to the structure of the antiferromagnetic domains.

Our samples are 30-nm- and 50-nm-thick CuMnAs films grown by molecular beam epitaxy on a GaP (001) substrate (Methods). Tetragonal CuMnAs films consist of alternating layers with opposite in-plane magnetization. The crystal and magnetic structure of CuMnAs are shown in Fig.~\ref{fig1}a. The possible set of domain orientations in these samples is restricted by the magnetic anisotropy: in thinner films ($t < 50\unit{nm}$), the anisotropy tends to be uniaxial with a $180^\circ$ reorientation of the N\'eel vector between adjacent domains. Thicker films have a stronger biaxial component and both $90^\circ$ and $180^\circ$ domain walls are present \cite{Wadley2018,Wadley2015a}. XMLD-PEEM images indicate N\'eel-type domain walls with the N\'eel vector rotating in the $a-b$~plane \cite{Wadley2018}.

Figure~\ref{fig1}b depicts a schematic of the measurement principle. A diamond probe containing a single NV centre at the apex is scanned at constant height ($z\sim 50-100\unit{nm}$) above the sample surface.  At every location we measure the shift in the NV centre's spin resonance using optical readout \cite{Balasubramanian2008}, which is directly proportional to the magnetic stray field $\BNV(x,y)$ at that location.  Note that the NV centre is sensitive only to the field component parallel to its symmetry axis, which lies at an angle $\theta=55^\circ$ off the surface normal for our probes (see inset to Fig.~\ref{fig1}b).  Using the known vector orientation of the NV center, the full magnetic vector field at the NV centre position can be reconstructed. Figure~\ref{fig1}c shows an example magnetic stray field map recorded from a 30~nm-thick CuMnAs film.

To recover the domain pattern and analyze the measurements, we model the antiferromagnet by two thin layers of opposite polarization located at the top and bottom of the film, respectively (Supplementary Section S1). Each layer carries a surface magnetization of $\Ms = n m s/V$, where $n$ is the number of \Mn ions per polarity per unit cell, $m$ is the magnetic moment per \Mn ion, $s=0.2$~nm is the vertical separation between oppositely polarized \Mn ions, and $V$ is the unit cell volume. The total magnetic stray field measured by the NV center is then given by the sum of top and bottom contributions, and dominated by the top layer because of the closer proximity to the probe.

In general, it is not possible to unambiguously reconstruct the magnetization from the magnetic stray field map, as the divergence-free part of the magnetization does not produce a stray field outside of the material \cite{Beardsley1989}. However, a rigorous reconstruction of the domain pattern can be obtained if we assume (i) uniaxial domain orientation with $|\Ms(x,y)|=\Ms$, and (ii) neglect the finite width of the domain walls (see Supplementary Section S2). Condition (i) can be assumed for thinner samples ($t<50$~nm) which have an uniaxial anisotropy\cite{Wadley2015a}, whereas (ii) is an approximation that does not affect the conclusions of our study. The reconstructed domain pattern from Fig.~\ref{fig1}c is shown in Fig.~\ref{fig1}d. We find that the easy axis is approximately pointing along the [110] ($\pm y$) direction. Note that the magnetic field map in Fig.~\ref{fig1}c reflects  the morphology of the domain pattern shown in Fig.~\ref{fig1}d, provided that the spatial sensitivity is comparable to the feature size. Comparison with XMLD-PEEM images \cite{Grzybowski2017,Wadley2018} reveals similar domain patterns as those observed in Fig.~\ref{fig1}d. Additional control measurements and simulations exclude ferromagnetic defects as the origin of the magnetic signals as these give rise to a much larger stray field (see Supplementary Section S3).

\subsection*{Current distribution and switching of CuMnAs microdevices}
To investigate the effect of an electric current on the domain pattern, we combine electrical resistance measurements and scanning NV magnetometry. Figure~\ref{fig2} shows the cross-shaped geometry of a patterned CuMnAs device used for electrical pulsing experiments. According to previous XMLD-PEEM studies \cite{Grzybowski2017,Wadley2018}, cross shaped devices enable 90$^{\circ}$ switching of the N\'eel vector by applying orthogonal current pulses or by flipping the polarity of the pulses. We define the orthogonal current directions
$P0^\pm$ and $P1^\pm$ in Fig.~\ref{fig2}a, where $\pm$ indicates the polarity of the pulse.

In a first step, we image the current density distribution by recording the Oersted field \cite{chang2017nanoscale} (see Supplementary Section S4). In Fig.~\ref{fig2}b we show the current distribution for the current direction $P0^\pm$. The current density is highest at the corners of the cross, as expected, and presents a granular texture that changes from device to device. Figure~2c shows an analogous current density map recorded for $P1^\pm$.

The magnetic stray field map in Fig.~\ref{fig2}e shows the pristine state of the domain pattern of the 50~nm thick CuMnAs film in the lower corner of the cross (dashed square in Fig.~\ref{fig2}b), before any current pulses are applied. After applying a $P0^+$ pulse of 56~mA and duration 100~$\mu$s, corresponding to an average current density $J=1.58\times10^7$~A~cm$^{-2}$ over the nominal $\sqrt{2}$(50~nm $\times$ 5~$\mu$m) cross-section of the device, we record a second magnetic stray field map (Fig.~\ref{fig2}f). When comparing Figs.~2e and 2f, we observe no change of the stray field pattern due to a
$P0^+$ pulse. This observation is consistent with the current flowing mostly in the adjoining corners of the cross and a negligible current density in the mapped region.
A $P1^+$ pulse, on the other hand, produces a high current density in the imaged area (Fig.~\ref{fig2}c), and we observe a significant change in the domain pattern. The strongest changes occur in the lower portion of the scan, where the current density is highest. This is consistent with the expected switching of the N\'eel vector in regions of high current density. Inverting the polarity of the pulse to $P1^-$ also leads to a change of the domain pattern, as seen in Fig.~\ref{fig2}h. The variation of the stray field between Figs.~2g and 2h is compatible with 180$^\circ$ switching of the N\'eel vector, as suggested by previous electrical measurements for pulses of opposite polarity \cite{Godinho2018}, possibly in combination with 90$^\circ$ switching due to domain wall motion \cite{Wadley2018}.

In addition to mapping the stray field and current density, we also record the transverse resistance
$\Rxy$ by sending a readout current along one arm of the cross and measuring the voltage perpendicular to the current flow \cite{Wadley2016,Grzybowski2017}. The readout current amplitude is about fifty times weaker than the writing current and safely below the switching threshold. Electrical measurements performed after $P0$ ($P1$) pulses show a decrease (increase) of $\Rxy$ of $\sim 160\unit{m\Ohm}$ with a characteristic decay time in the tens of seconds (Fig.~\ref{fig2}d). $\Rxy$ is recorded for every pixel in the NV images simultaneously with $\BNV$. The time required to acquire a complete $\BNV(x,y)$ map is about 3~h. The NV maps, therefore, correspond to a relaxed $\Rxy$ signal (grey regions in Fig.~\ref{fig2}d). Despite an almost complete relaxation of the $\Rxy$ signal, a tiny $\sim 5\unit{m\Ohm}$ difference remains between the relaxed $\Rxy$ values for $P0$ and $P1$ pulses even after many hours (inset to Fig. \ref{fig2}d), corresponding to a resistive switching ratio of less than 0.1\% relative to the $20\unit{\Ohm}$ sheet resistance of our film. Similarly weak electrical switching signals were linked in previous XMLD-PEEM studies to the 90$^\circ$ reorientation of the N\'eel vector in the antiferromagnetic domains and to the corresponding AMR \cite{Wadley2016,Grzybowski2017}. The initial values of $\Rxy$ before relaxation are, however, two orders of magnitude higher than the relaxed values. Both the large amplitude and the relaxation in a tens of seconds time-scale at room temperature have been reported in recent electrical and optical pulsing experiments \cite{Kaspar2019}. These observations point to a new switching mechanism unrelated to the net N\'eel vector reorientation and AMR. In the following section we focus on the NV imaging of the antiferromagnetic state corresponding to these initial unrelaxed $\Rxy$ signals within a few seconds after the writing pulse.

\subsection*{Current-induced domain fragmentation}

To probe the domain structure during the initial fast decay of the electrical resistance, corresponding to the grey shaded region in Fig.~\ref{fig3}a, we implement a pump-probe scheme that interleaves the data acquisition with electrical current pulsing. We also vary the current density to probe the relaxation after pulses that induce changes of $\Rxy$ in the m$\Omega$ to $\Omega$ range, as shown in Fig.~\ref{fig3}b. In the pump-probe method, described in Fig.~\ref{fig3}c, we apply a writing current pulse before the acquisition of each pixel and measure $\BNV$ during the first 4~s right after the pulse. Two scans are recorded at the same time, with their pixels interleaved, one after application of a $P0^+$ pulse, the other after a $P1^+$ pulse. In this way we are able to probe the magnetic state averaged over the first 4~s of the relaxation process for both current directions. For these measurements we probe the central region of the cross shaped devices, where approximately the same current density can be expected for $P0$ and $P1$ pulses. Examples of stray field images acquired in the pump-probe mode for a 30~nm thick CuMnAs film are shown in Figs.~3d-g.

We first focus on writing currents close to the density threshold of the large switching $\Rxy$ signal. In Fig.~\ref{fig3}a we show the time-dependence of $\Rxy$ after a 100~$\mu$s writing pulse of average current density $J=1.89\times10^7$~A~cm$^{-2}$, which is just above the threshold shown in Fig.~\ref{fig3}b.
The first image (Fig.~\ref{fig3}d) acquired after $P0^+$ pulses with $J=1.89\times10^7$~A~cm$^{-2}$ shows a stray field pattern very similar to that of the pristine sample, indicating that the current density in the centre of the cross is not sufficient to modify the antiferromagnetic domains in an appreciable way. Upon increasing the current density to $J=1.98\times10^7$~A~cm$^{-2}$, however, we observe a striking reduction of the amplitude of the magnetic stray field (Fig.~\ref{fig3}f), which we quantify by taking the root mean square of $\BNV(x,y)$ over the entire magnetic field map, $B_\text{rms}$, see methods. The reduction in $B_\text{rms}$ is similar for the images acquired after $P0^+$ and $P1^+$ pulses (Fig.~\ref{fig3}e,f), indicating that the direction of the writing current does not play a role in this effect.
%

Once the pulsing stops, the stray field amplitude 
slowly recover on a time scale of days. The relaxed image (Fig.~\ref{fig3}g), which is acquired 75~h after applying the last pulse from Fig.~\ref{fig3}f, shows that the system maintains a memory of the pristine domain configuration even after a long sequence of excitations. The memory effect is not perfect, as can be seen by comparing the upper left corner of Fig.~\ref{fig3}d-g, but pervasive to both the excited and relaxed states.

We now argue that the reduction of the stray field amplitude is caused by a decrease of the average domain size.
This behavior can be understood by simulating the stray field produced by varying domain configurations.
Starting from a model of the pristine domain configuration and its stray field pattern, shown in Figs.~\ref{fig4}a and \ref{fig4}b, respectively, we generate a magnetization pattern with a fragmented structure, retaining the overall shape of the domain pattern as defined by regions with a prevailing orientation of the N\'eel vector (Figs.~\ref{fig4}c,d and Supplementary Section S5).
The simulated stray field maps of the pristine and fragmented domains present a similar morphology but a different magnetic contrast, in agreement with the experiment.
Alternative explanations to the reduction in magnetic contrast, such as heat-induced suppression of the magnetization $\Ms$ or a change in sensor stand-off distance $z$ can be safely excluded, as the relaxation occurs on a much longer time scale compared to thermal effects and no drifts in the scanning setup are observed.

The decrease of $\Brms$ can be qualitatively understood by noting that $\BNV$ at a height $z$ above the surface is most affected by changes of the magnetization that occur on the same length scale as $z$. Much larger and homogeneous structures generate stray fields only in the vicinity of the domain walls, whereas stray field lines of more localized structures are confined to the close proximity of the surface.
In the context of domain imaging, this means that finely broken-up domains are too small to be resolved by the NV sensor.
Comparing the reduction in $\Brms$ to the results of our numerical simulations (Fig.~\ref{fig4}e), we estimate that the current pulsing leads to the formation of fragmented domains with a typical length scale of about 10~nm.

\subsection*{Correlation between domain fragmentation and resistive readout signal}
To gain further insight into the relationship between magnetic domains and electrical resistance, we performed a series of pump probe measurements for only one current polarity as a function of pulse amplitude. Figure~\ref{fig5}a plots the transverse resistance $\Rxy$ as a function of time of a 50-nm-thick CuMnAs device while the current density is stepped up from $1.36$ to $1.54\unit{A cm^{-2}}$. The measurements are performed in the lower corner of the cross, where the impact of the current is highest (see Fig.~\ref{fig2}). At each current density step, we record a stray field map using the pump-probe scheme of Fig.~\ref{fig3} and compute $\Brms$. 
As expected, the resistance signal increases with increasing current density. In addition, we observe that repeated pulsing at one set value of $J$ leads to a further gradual increase of $\Rxy$. Figure~\ref{fig5}b shows that the increase in resistance is accompanied by a similar reduction in $\Brms$, clearly showing the correlation between the two effects.  The correlation persists after the pulsing stops and the system slowly evolves towards the relaxed state (hours 38-60).

This series of measurements demonstrates that the reduction in $\Brms$ 
does not depend on either the direction or polarity of the current, since it occurs for both orthogonal (Fig.~\ref{fig3}) and unidirectional pulses (Fig.~\ref{fig5}).  Further experiments involving bipolar pulses show that the fragmentation occurs in combination with domain switching, as fragmented and relaxed images also show evidence of 180$^\circ$ reorientation of the N\'eel vector in certain areas of the scans (see Supplementary Section S6).


\subsection*{Discussion and outlook}
The fragmentation of the domain pattern in CuMnAs by electrical pulses and the subsequent recovery is distinct from the reorientation of the N\'eel vector in the antiferromagnetic domains demonstrated in the past.  The fragmentation of the domains, as measured by the vanishing stray field contrast, 
increases with the number of current pulses and current density. Regions of highly fragmented domains are distributed non-uniformly throughout the sample, following the inhomogeneous current density distribution evidenced by Oersted field maps (see Supplementary Section S6). Fragmentation occurs for different pulsing strategies, namely for orthogonal pulses, reversed-polarity pulses, and unidirectional pulses, and in samples of different thicknesses (30~nm or 50~nm). We surmise it is a general effect likely governed by current-induced heating. The degree of domain fragmentation correlates with the increase of the resistive readout signal up to high amplitudes that are well above the earlier identified AMR signals due to the reorientation of the N\'eel vector in the antiferromagnetic domains. The domain fragmentation thus provides a plausible explanation for the recently observed unipolar high-resistive switching signals in CuMnAs and their relaxation\cite{Kaspar2019}.

However, several points remain open for discussion. First, the influence of the fragmented state on the electrical resistance might be explained by the formation of a dense network of antiferromagnetic domain walls. In ferromagnets, the influence of domain walls on the electrical resistivity is well documented \cite{Berger1978,Levy1997,Tang2004}.  For example, the striped domain phase of a Co film with a density of approximately 0.005~nm$^{-1}$ domain walls perpendicular to the current flow causes a 5\% increase of the resistivity compared to the uniform magnetic state \cite{Letters1996}. In CuMnAs, the relative change of the resistivity reaches $\sim$100\% at low temperature \cite{Kaspar2019}. Such a large effect can be possibly explained by the very high density of domain walls attainable in antiferromagnets. Our spatial resolution points to a linear density of domain walls larger than 0.02~nm$^{-1}$, with the simulations indicating a change of density from 0.01~nm$^{-1}$ to 0.1~nm$^{-1}$ required to match the observed reduction of the magnetic stray field. Although the large density of domain walls is striking, little is known about the influence of domain walls on the electrical transport in antiferromagnets in general \cite{Jaramillo2007,Kummamuru2008}. Our findings might stimulate more work in this direction.

Second, it is unclear at this stage whether the magnetic fragmentation is accompanied by a modification of the crystalline structure of the antiferromagnet and perhaps driven by it. Surface sensitive photoemission, electron and atomic force microscopy as well as bulk-sensitive transmission electron microscopy do not show  structural changes in CuMnAs for pulsing conditions at which the highly reproducible resistive switching signals are observed, provided that they stay safely below the device breakdown threshold \cite{Krizek2019}.

Third, the memory effect of the antiferromagnetic domain pattern is quite striking in itself. The samples that have undergone domain fragmentation relax to a stray field pattern that is very similar in intensity and spatial features to that of the domain configuration prior to pulsing. In general, only small scan areas show evidence of permanent switching, which we attribute to
90$^\circ$ and 180$^\circ$ reorientation of the antiferromagnetic moments by current-induced torques, consistent with earlier XMLD-PEEM studies\cite{Grzybowski2017}. The memory of the domain pattern might point to a significant role of static defects in the film which could assist the magnetic effect as domain wall nucleation and pinning sites (an extensive characterization of defects in the epitaxial single-crystal films of CuMnAs is discussed elsewhere \cite{Krizek2019}). While still not fully understood microscopically, the possibility to switch between two distinct antiferromagnetic  phases while retaining a memory of the pristine state demonstrates a new type of memristor in which information is encoded in the magnetic structure rather than in the crystal structure.

In conclusion, scanning NV magnetometry measurements revealed the complexity of the excited antiferromagnetic texture after current injection, showed a correlation between domain fragmentation and  large resistive switching signals, and demonstrated a novel memristive effect. A full understanding of the mechanisms behind domain fragmentation, relaxation, and memory of the pristine domain configuration will require simultaneous investigation of the magnetic state and crystal structure on a local scale with bulk sensitivity and covering a broad range of time scales.

\subsection*{Acknowledgments}
This work was funded by the Swiss National Science Foundation (Grant No. 200020-172775), by the Swiss Competence Centre for Materials Science and Technology (CCMX).
C.L.D acknowledges funding from the Swiss National Science Foundation (SNFS) through Project Grant No. 200020\_175600 and the National Center of Competence in Research in Quantum Science and Technology (NCCR QSIT), from the Advancing Science and TEchnology thRough dIamond Quantum Sensing (ASTERQIS) program, Grant No. 820394, of the European Commission, and from ERC CoG ``IMAGINE'', Grant No. 817720, from the European Research Council. V.N. acknowledges funding from the Ministry of Education of the Czech Republic Grants LM2015087 and LNSM-LNSpin. T.J. acknowledges funding from the Czech National Science Foundation Grant No. 19-28375X and the EU FET Open RIA Grant No. 766566.

\subsection*{Author contributions}
M.S.W., P.W., T.J., C.L.D., and P.G. conceived the work and designed the experiments; Z.K., K.O., and V.N. fabricated the samples. M.S.W. and P.W. built the NV microscope and performed the measurements, which were interpreted by M.S.W., Z.K., K.O., T.J., C.L.D., and P.G. The reconstruction of the magnetic stray field maps and simulations were performed by M.S.W. and C.L.D.. M.S.W., T.J., C.L.D., and P.G. co-wrote the manuscript. All authors discussed the data and commented on the manuscript.

\subsection*{Additional information}
Correspondence and requests for materials should be addressed to\\
M.S.W. (\verb"martin.woernle@mat.ethz.ch"), C.L.D. (\verb"degenc@ethz.ch"), \\
and P.G. (\verb"pietro.gambardella@mat.ethz.ch").

\subsection*{Methods}

{\bf Samples:} Tetragonal CuMnAs films were grown by molecular beam epitaxy at 210$^\circ$C on GaP(001) substrates \cite{Krizek2019}. The films were protected by a 3-nm-thick Al capping layer to prevent oxidation of CuMnAs.
The samples were patterned into four-arm cross-shaped devices using electron beam lithography and wet chemical etching.

\begin{figure}[h]
\hspace*{-0cm}\includegraphics[trim={2cm 8cm 11cm 2cm},width=.9\textwidth]{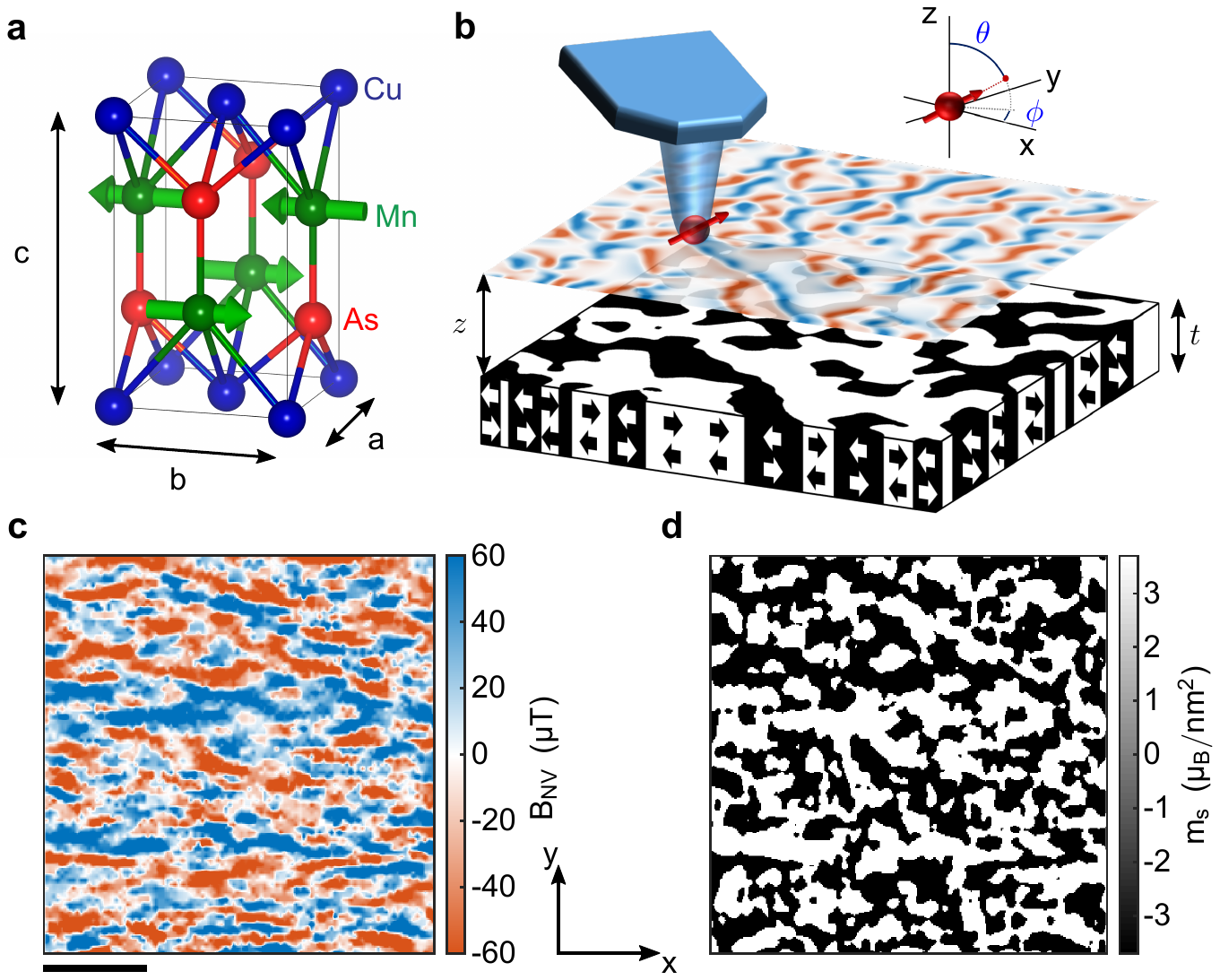}
\caption{{\bf Scanning NV magnetometry on CuMnAs.}
		{\bf a,} Unit cell of CuMnAs. The magnetic moments of the \Mn ions (green arrows) are oriented in plane and alternate along the $[001]$ direction (the $c$-axis).
		{\bf b,} Schematic of the scanning NV magnetometer.  A diamond tip (blue) containing an NV centre (red arrow) is scanned over an antiferromagnetic film (thickness $t=30-50\unit{nm}$).  Antiferromagnetic domains are represented by black and white areas with co-planar spins.  The scanning NV magnetometer records the antiferromagnet's magnetic stray field $\BNV(x,y)$ at a distance $z=50-100\unit{nm}$ above the surface (red/blue pattern).
		The inset defines the ($\theta$,$\phi$) vector orientation of the NV centre.
		{\bf c,} Example of a magnetic stray field map of a pristine 30-nm-thick CuMnAs film.  NV centre parameters are ($z=60\pm7$~nm, $\phi=270^\circ\pm5^\circ$, $\theta=55^\circ$).
		{\bf d,}  Domain pattern reconstructed from the field map in panel c, as described in the text.
		Scale bar, 800\,nm.
		}
\label{fig1}
\end{figure}

\begin{figure}[h]
\includegraphics[trim={4cm 7cm 6cm 0},width=.8\textwidth]{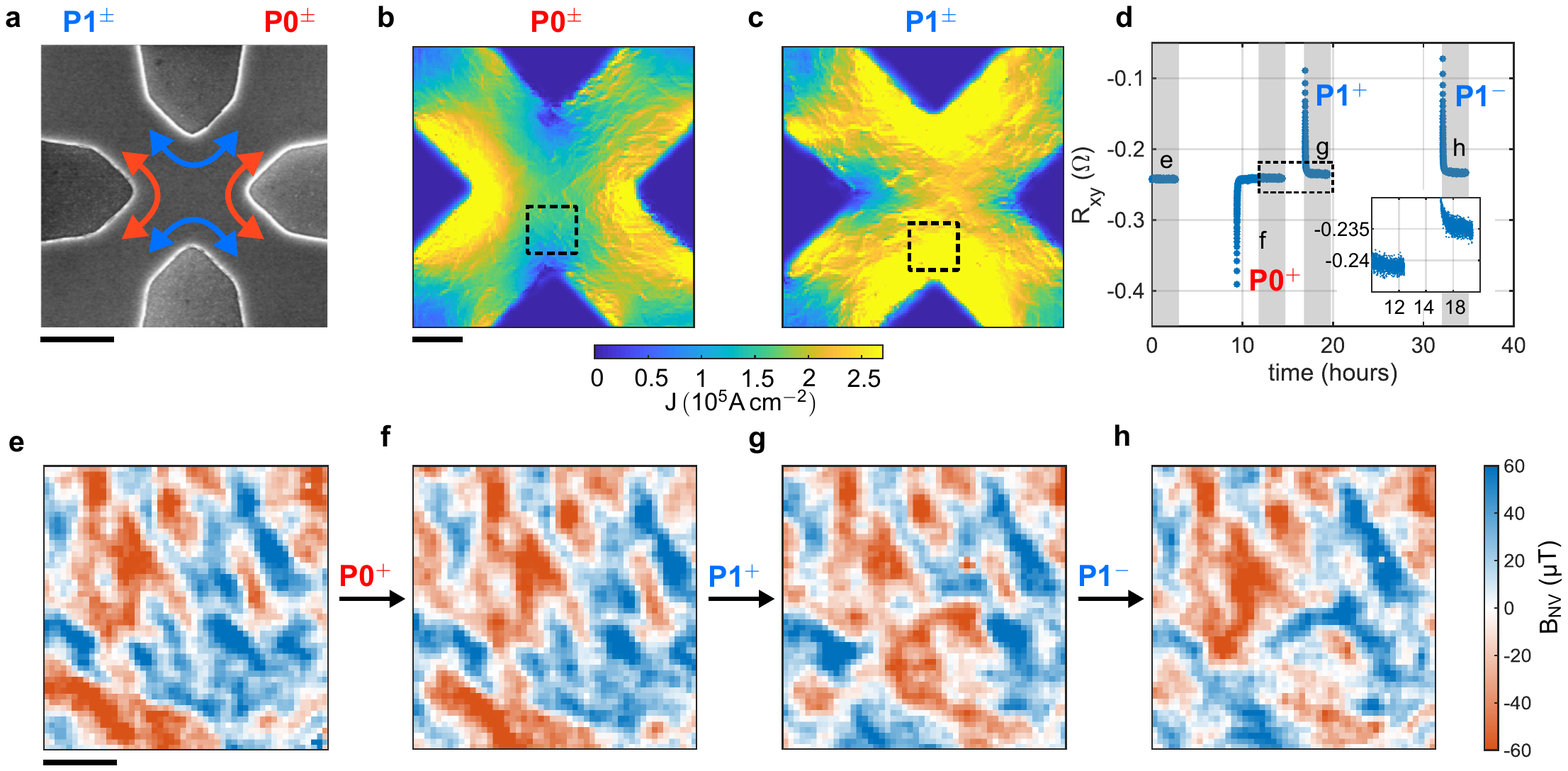}
\caption{{\bf Current distribution, electrical resistance, and magnetic stray field maps of the relaxed state after switching.}
		{\bf a,} Scanning electron micrograph of a representative cross-shaped CuMnAs device. The arms of the cross are 5-$\mu$m-wide and oriented parallel to the [100] and [010] crystal axes of the CuMnAs film. The pulse directions are defined as $P0^\pm$ (red arrows) and $P1^\pm$ (blue arrows). The sign denotes polarity. Scale bar, 5~$\mu$m.
		{\bf b,c,} Current density distribution for $P1$ and $P0$ pulses measured using scanning NV magnetometry for a probe current of 1~mA (see Supplementary Section S4). Scale bar, 2~$\mu$m.
		{\bf d,} Transverse electrical resistance $\Rxy$ as a function of time. Three current pulses of amplitude $J=1.58\times10^7$~A~cm$^{-2}$, duration 100~$\mu$s, and different direction are injected during this measurement, as indicated before each spike of $\Rxy$. The measurement are performed simultaneously with the stray field scans shown in panels e-h. The total acquisition time of each scan is indicated by the shaded grey regions. The inset magnifies the electrical measurement within the dashed rectangle. A difference in the electrical signal is still visible after hours.
		{\bf e-h}, Magnetic stray field images of the lower region of the cross (dashed square in panel b,c) before any pulsing (panel e), after a $P0^+$ pulse (f), after a $P1^+$ pulse (g), and after a $P1^-$ pulse (h). The NV sensor parameters are ($z=63\pm4$~nm, $\phi=185^\circ\pm5^\circ$, $\theta=55^\circ$).  Scale bar, 400\,nm.
		}
\label{fig2}
\end{figure}

\begin{figure}[h]
\hspace*{-0cm}\includegraphics[trim={5cm 7cm 6cm 2cm},width=.8\textwidth]{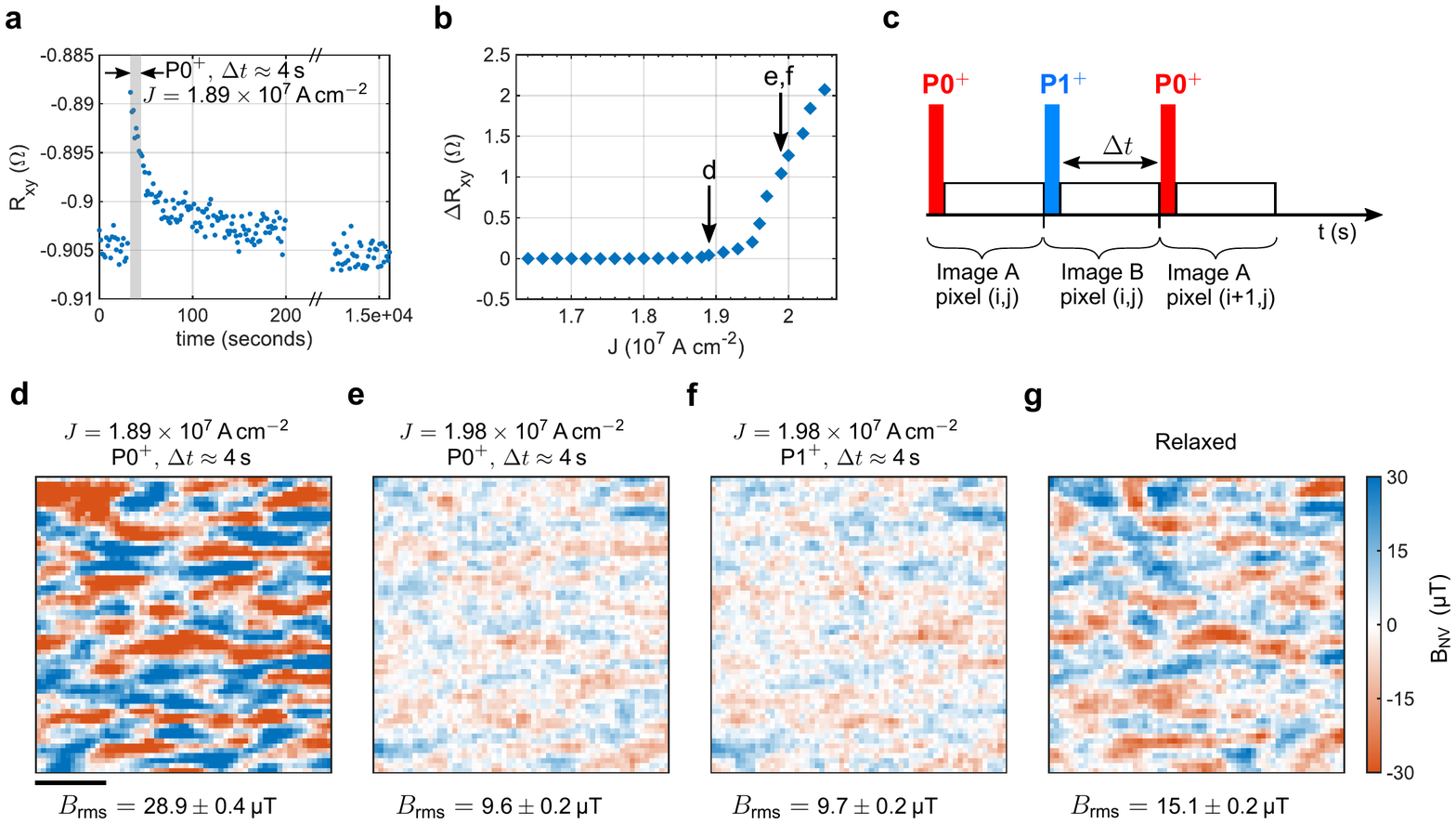}
\caption
{{\bf Pump-probe measurement scheme and stray field maps of the excited and relaxed states. } {\bf a,} Temporal evolution of $\Rxy$ after application of a single $P1^+$ current pulse of amplitude $J=1.89\times10^7$~A~cm$^{-2}$ and duration 100~$\mu$s to a 30~nm thick CuMnAs device. {\bf b,} Current density $J$ vs. maximum switching amplitude $\Rxy$. {\bf c,} Schematic of the measurement sequence. For each image pixel $(i,j)$ we measure the magnetic stray field twice, once after a $P0^+$ pulse and once after a $P1^+$ pulse. The stray field measurement starts immediately after a pulse and is integrated over $\Delta t=4\unit{s}$ (grey shaded area region in panel a).  This sequence is repeated pixel by pixel to build up the images shown in panels d-g. {\bf d-g,} Magnetic stray field maps of the 30~nm thick CuMnAs film after $P0^+$ pulses (d,e), $P1^+$ pulses (f), and 75 hours after the last pulse (g). The measurements are performed in the centre of the cross and the pulse amplitude is given above each scan. The sensor parameters are ($z\approx52\pm11$~nm, $\phi\approx88^\circ\pm5^\circ$, $\theta\approx55^\circ$). Scale bar, 400\,nm.
}	
\label{fig3}
\end{figure}

\begin{figure}[h]
\hspace*{-0cm}\includegraphics[trim={2cm 8cm 8cm 0cm},width=.8\textwidth]{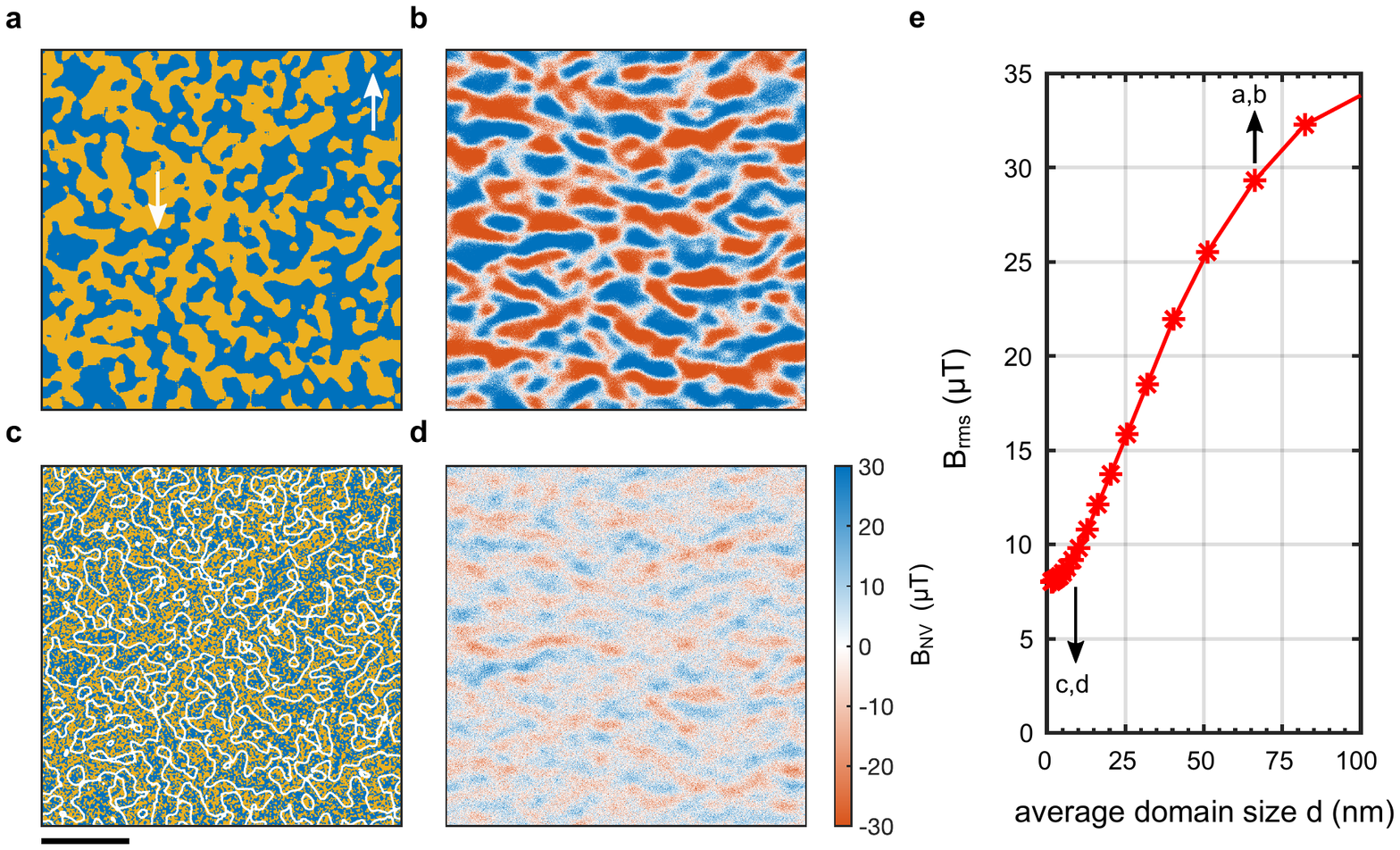}
	\caption{{\bf Simulations of the stray field produced by a pristine and a fragmented domain pattern.}
	{\bf a,b}, Simulated pristine domain configuration (panel a) and magnetic stray field (panel b) for a 30-nm-thick CuMnAs film.  Arrows indicate the direction of the N\'{e}el vector.
	{\bf c,d}, Simulated fragmented domain configuration (panel c) and magnetic stray field (panel d).  White contour lines mark the pristine domain walls from panel a. The fragmentation leads to a reduction of the magnetic contrast $B_\text{rms}$, whereas the overall shape of the stray field pattern is partially conserved. Scale bar, 400\,nm.
	{\bf e,} Simulated $\Brms$ as a function of average domain size $d$ defined as (number of domain walls per unit length)$^{-1}$.
	For small domains $d<z$, the stray field is approximately proportional to $d$.
	Black arrows indicate the approximate $d$ for the patterns plotted in panels a-d.
		Simulations use the same NV centre parameters as in Fig. \ref{fig3}. Gaussian noise  $(\SI{8}{\micro\tesla} - \text{rms})$ is added to the field maps to account for measurement noise. See Supplementary Section S5 for more details.
	}
\label{fig4}
\end{figure}

\begin{figure}[h]
\hspace*{-0cm}\includegraphics[trim={4.5cm 6cm 6cm 0},width=.8\textwidth]{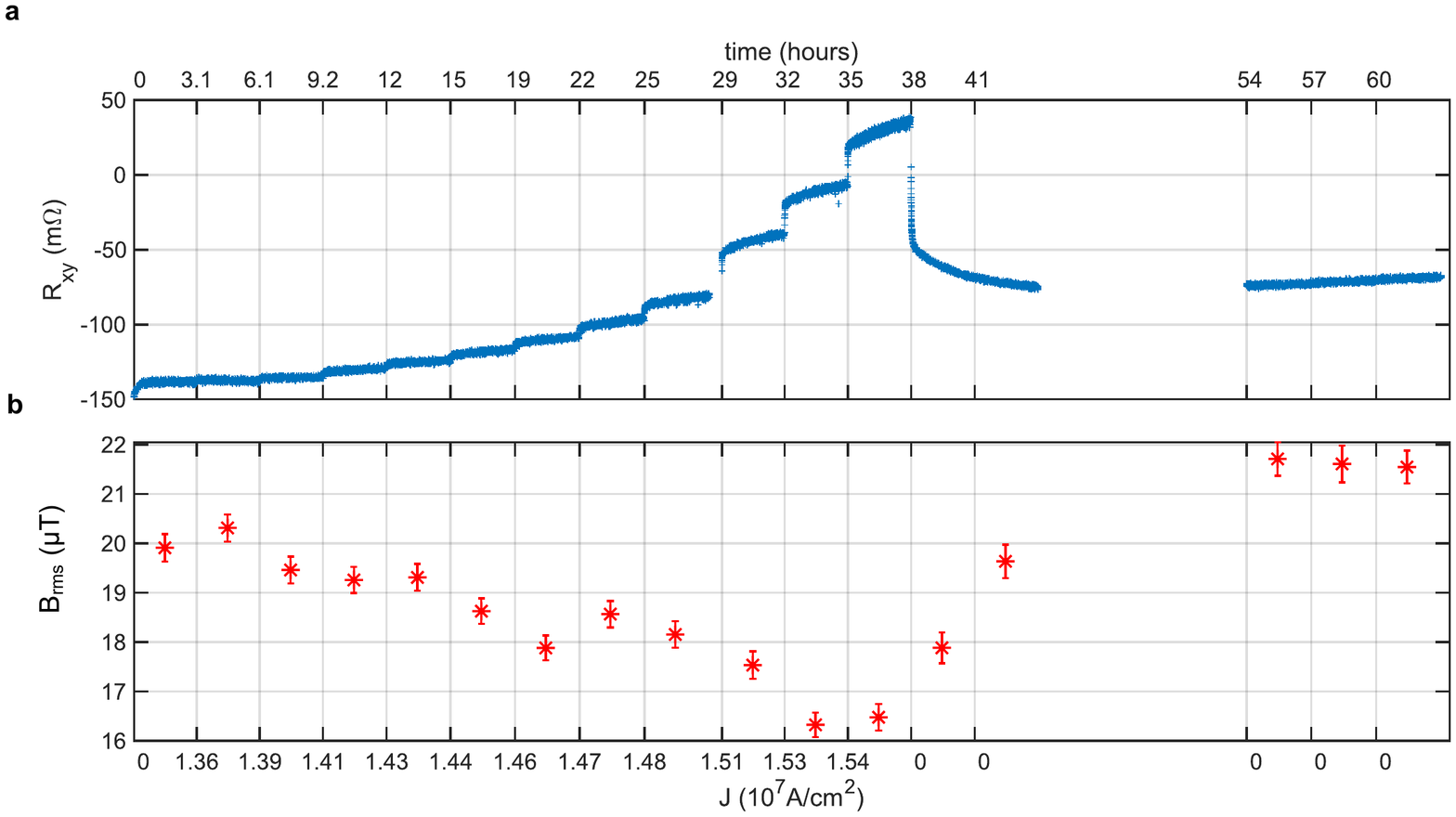}
\caption{{\bf Correlation between fragmentation and electrical resistance.}
		{\bf a,} Change of the transverse resistance $\Rxy$ and
		{\bf b,} Stray field amplitude $\Brms$ 
 versus current density $J$ (bottom axis) and time (top axis) of the 50~nm-thick CuMnAs film.  $\Rxy$ is plotted for every pixel in chronological order.  $\Brms$ is computed from stray field scans recorded at each $J$ value in 3-hour intervals (vertical lines).
The measurements are performed using the same pump-probe scheme as in Fig.~\ref{fig3}a, but only $P1^+$ pulses are applied.  The scan area is the same as in Fig.~\ref{fig2}b,c.  The NV sensor parameters are ($z=97\pm2$~nm, $\phi=96^\circ\pm3^\circ$, $\theta=55^\circ$).}
\label{fig5}
\end{figure}

{\bf NV magnetometry:}
To investigate the domain pattern of CuMnAs, we use scanning NV magnetometry to image the local magnetic stray field $B_\text{NV}\left(x,y\right)$ emanating from the domain walls, between adjacent domains.  The scans are carried out on a nanoscale scanning NV magnetometer microscope in ambient conditions of own design, built in-house. The microscope employs  monolithic diamond probe tips with single NV centres implanted at the apex (QZabre LLC, https://qzabre.com/). By optically-detected magnetic resonance (ODMR) spectroscopy\cite{Gruber1997,Schirhagl2014} using a nearby microwave antenna ($2.9$~GHz) and optical (532~nm excitation, 630-800~nm detection) readout we monitor the NV center spin resonance. We apply a small magnetic bias field (2-4~mT) to obtain a sign-sensitive measurement of the magnetic stray field. The spin resonance is obtained by fitting a Lorentzian to the ODMR spectrum. We convert this resonance frequency to units of magnetic field by looking at the resonance frequency shift $\Delta f$, referenced on the resonance frequency when far away from the sample. The detected field $\BNV$ is then given by $\BNV=2\pi \Delta f/\gamma$, where  $\gamma=2\pi\cdot28.0$~GHz/T is the electron gyromagnetic ratio. By knowing the sign of the bias field the sign of the magnetic stray field can be deduced. Note that the NV centre is sensitive only to fields that are parallel to its symmetry axis, determined by the crystallographic orientation of the diamond tip and the probe arrangement in the setup. The measured $\BNV$ therefore represents the projection of the vector field $\mathbf{B}=(B_x,B_y,B_z)$ onto the NV center spin's symmetry axis,
$\BNV = B_x \sin\theta \cos\phi + B_y \sin\theta \sin\phi + B_z \sin\theta$,
where $\theta$ and $\phi$ are the polar and azimuth angle of the NV spin symmetry axis in the laboratory frame of reference (see inset to Fig. \ref{fig1}b).
The $\theta,\phi$ angles are calibrated beforehand by a series of ODMR measurements in different magnetic bias fields, and confirmed by line scans.
The sample-to-sensor distance $z$ is inferred on an out-of-plane magnetized $\text{Pt/Co/AlO$_x$}$ stripe.
More details on the calibration of $\theta$, $\phi$ and $z$ are given in Supplementary Section S7.
The magnetic stray field patterns shown in Figs.~1-3 have been low-pass filtered with a Gaussian filter ($\sigma = 24\unit{nm}$) to better highlight the morphology.

We quantify the amplitude of the magnetic stray field pattern by taking the root mean square of $\BNV(x,y)$ over the entire magnetic field map
\begin{align}
	\Brms = \sqrt{\frac{1}{n}\sum_{i=1}^n (\BNV^i)^2},
\end{align}
where the index $i$ enumerates the pixels.
The uncertainty of $B_\text{rms}$ is estimated via jackknife resampling\cite{efron1981jackknife}. For this, we repeatedly compute the RMS value, each time leaving out one single point from the dataset. The uncertainty can then be inferred from the spread of these values. Let $x_i$ be the RMS value when omitting the $i$-the data point, and let $x$ be their mean. The estimated uncertainty of the RMS of the complete dataset is
\begin{align}
	\Delta\Brms  = \sqrt{\frac{n-1}{n}\sum_{i=1}^n (x_i - x)^2}.
\end{align}
%




%


\clearpage

\clearpage

\clearpage

\clearpage

\clearpage

\end{document}